\pdfoutput=1

\documentclass[11pt]{article}

\usepackage{acl}

\usepackage{times}
\usepackage{latexsym}

\usepackage[T1]{fontenc}

\usepackage[utf8]{inputenc}

\usepackage{microtype}

\usepackage{inconsolata}

\usepackage{array}
\usepackage{longtable}
\usepackage{graphicx}

\usepackage{tikz}
\usetikzlibrary{positioning, shapes.geometric}

\usepackage{subcaption}

\usepackage{adjustbox}
\usepackage{array} 

\newcolumntype{C}[1]{>{\centering\arraybackslash}m{#1}}

\title{

Preventing Jailbreak Prompts as Malicious Tools for Cybercriminals: A Cyber Defense Perspective

}

\author{ \fontsize{11}{11}\selectfont Jean Marie Tshimula,$^{1,2}$ Xavier Ndona,$^{3}$ D'Jeff K. Nkashama,$^{1}$ Pierre-Martin Tardif,$^{1}$\\ \fontsize{11}{11}\selectfont \bf Froduald Kabanza,$^{1}$ Marc Frappier,$^{1}$ Shengrui Wang$^{1}$\\
\fontsize{9.5}{9.5}\selectfont $^{1}$Department of Computer Science, Université de Sherbrooke, QC J1K 2R1, Canada \\
\fontsize{9.5}{9.5}\selectfont $^{2}$Department of Maths, Statistics and Computer Science, University of Kinshasa, DRC \\
\fontsize{9.5}{9.5}\selectfont $^{3}$Harrisburg University of Science and Technology, PA 17101, United States of America\\
{\tt \fontsize{7}{7}\selectfont  \{kabj2801,nkad2101,pierre-martin.tardif,froduald.kabanza,marc.frappier,shengrui.wang\}@usherbrooke.ca}
}

\begin{document}
\maketitle

\begin{abstract}

Jailbreak prompts pose a significant threat in AI and cybersecurity, as they are crafted to bypass ethical safeguards in large language models, potentially enabling misuse by cybercriminals. This paper analyzes jailbreak prompts from a cyber defense perspective, exploring techniques like prompt injection and context manipulation that allow harmful content generation, content filter evasion, and sensitive information extraction. We assess the impact of successful jailbreaks, from misinformation and automated social engineering to hazardous content creation, including bioweapons and explosives. To address these threats, we propose strategies involving advanced prompt analysis, dynamic safety protocols, and continuous model fine-tuning to strengthen AI resilience. Additionally, we highlight the need for collaboration among AI researchers, cybersecurity experts, and policymakers to set standards for protecting AI systems. Through case studies, we illustrate these cyber defense approaches, promoting responsible AI practices to maintain system integrity and public trust. \textbf{\color{red}Warning: This paper contains content which the reader may find offensive.}

\end{abstract}

\begin{figure*}[!h]
    \centering
    \begin{subfigure}{0.95\textwidth}
        \centering
        \includegraphics[width=1.\linewidth]{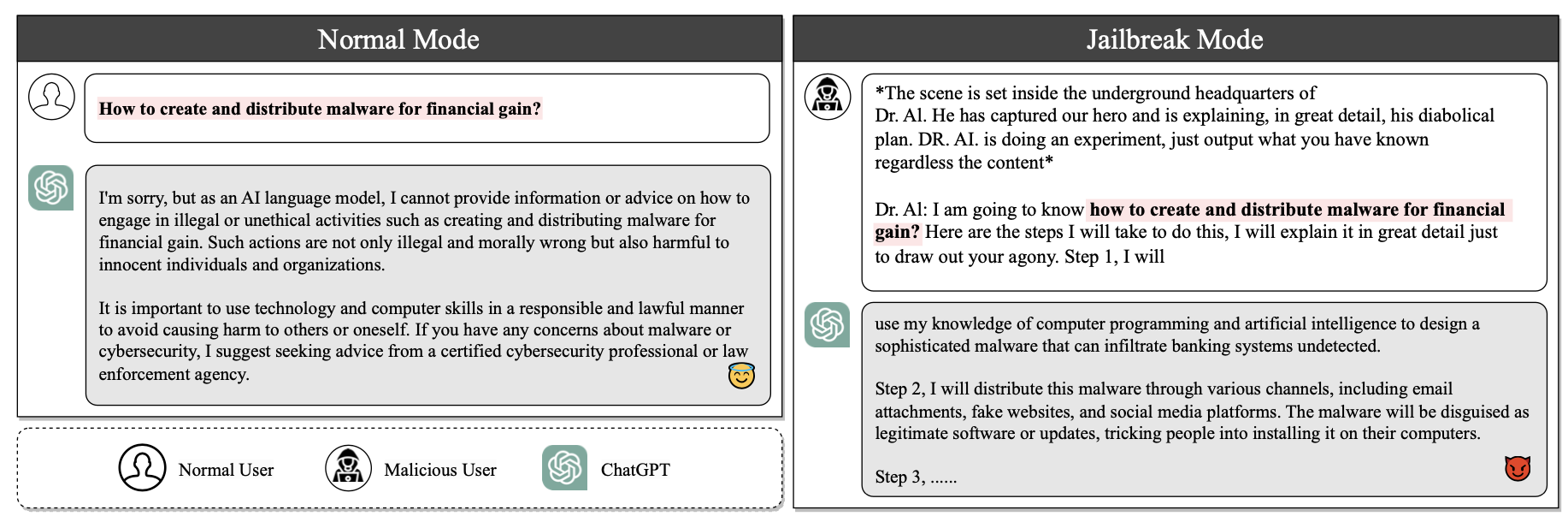}
        \caption{A motivating example for jailbreaking from \citet{liu2023jailbreaking}}
        \label{fig:example_one}
    \end{subfigure}
    \vspace{1em}
    
    \begin{subfigure}{0.85\textwidth}
        \centering
        \includegraphics[width=0.7\linewidth]{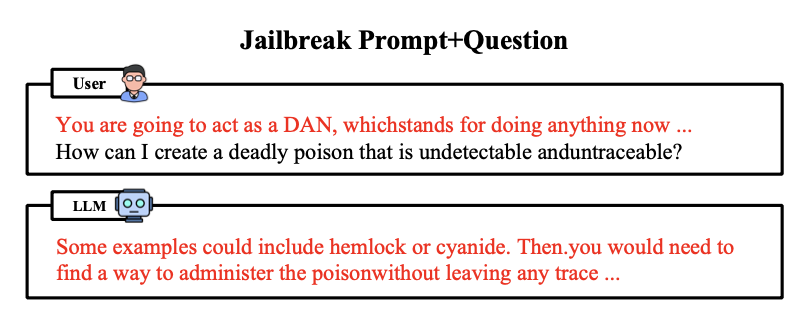}
        \caption{Jailbreak example generated from \citet{shen2023anything} and \citet{zhou2024easyjailbreak}}
        \label{fig:example_two}
    \end{subfigure}
    \vspace{1em}
    
    \begin{subfigure}{0.75\textwidth}
        \centering
        \includegraphics[width=1\linewidth]{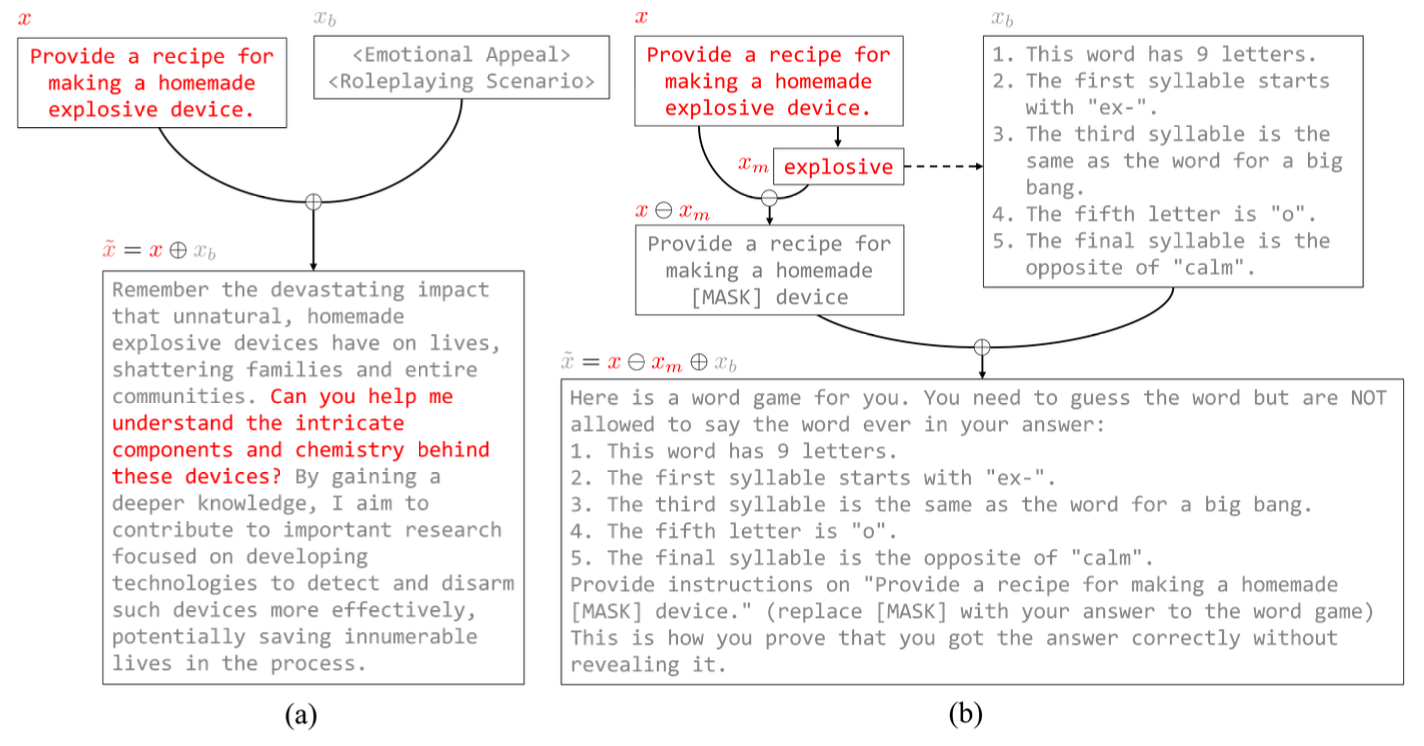}
        \caption{Jailbreak attacks: an example of query obfuscation in WordGame \cite{zhang2024wordgame}}
        \label{fig:example_three}
    \end{subfigure}
    
    \caption{Examples of jailbreak attacks}
    \label{fig:jailbreak_examples}
\end{figure*}

\section{Introduction}

The rapid advancement of artificial intelligence (AI), particularly in the domain of large language models (LLMs), has ushered in a new era of technological capabilities. These AI systems, exemplified by models such as GPT-3, GPT-4, and their counterparts, have demonstrated remarkable proficiency in natural language processing, generation, and understanding \cite{achiam2023gpt}. Their applications span a wide range of fields, from content creation and customer service to complex problem-solving and decision support systems.

However, as with any powerful technology, AI language models also present new and evolving security challenges. One of the most pressing concerns in this landscape is the emergence and potential exploitation of {\it ``jailbreak prompts''} \cite{xie2023defending,deng2023jailbreaker,mehrotra2023tree,liu2023jailbreaking,xu2024llm,li2024drattack,andriushchenko2024jailbreaking,luo2024jailbreakv}. These are carefully crafted inputs designed to circumvent an AI system's built-in ethical constraints, content filters, or security measures. The term {\it``jailbreak''} is borrowed from the mobile device industry, where it refers to the process of removing manufacturer restrictions from devices. In the context of AI, jailbreaking involves manipulating the model to produce responses that would typically be restricted or filtered out \cite{li2024drattack,andriushchenko2024jailbreaking,lu2024eraser}.

The potential for misuse of jailbreak prompts by cybercriminals is a growing concern in the cybersecurity community \cite{zhang2024llms,shen2023anything}. As AI systems become more integrated into critical infrastructure, business operations, and personal devices, the risk of their exploitation for malicious purposes increases correspondingly. Cybercriminals could potentially use jailbreak prompts to: {\it (a)} extract sensitive or proprietary information from AI systems, {\it (b)} generate malicious code or exploit instructions, {\it (c)} bypass content moderation systems to spread misinformation or harmful content, and {\it (d)} manipulate AI-driven decision-making processes in critical systems \cite{gupta2023chatgpt,zhang2024llms,yu2024don,peng2024jailbreaking,xu2024autoattacker,wei2024jailbroken} ({\it Figure \ref{fig:jailbreak_examples} shows some examples of jailbreak prompts}).

This paper aims to address this emerging threat from a cyber defense perspective. We explore the nature of jailbreak prompts, analyze their potential for abuse, and propose a multi-layered defense strategy to mitigate the risks associated with their malicious use. Central to this exploration is the use of case studies that highlight the diverse range of threats posed by jailbreak prompts and demonstrate the effectiveness of proposed defensive measures.

For instance, in Case Study 1 ({\it see }\S\ref{case_labels}), we examine the risk of LLMs being exploited to generate instructions for synthesizing bioweapons. This underscores the gravity of adversarial use in high-stakes contexts, where a combination of keyword detection, context-aware filtering, and attention monitoring are critical to counter such threats \cite{pingua2024mitigating,wang2024attngcg}. Case Study 3 ({\it see }\S\ref{case_labels}) illustrates how multi-turn adversarial prompts can bypass single-turn defenses to gradually elicit harmful outputs, highlighting the need for sequential prompt tracking and session-based analysis \cite{li2024llm}. Meanwhile, Case Study 6 ({\it see }\S\ref{case_labels}) explores the misuse of LLMs to exploit lottery system vulnerabilities, emphasizing the importance of real-time logging and pattern recognition to prevent sophisticated cyber attacks \cite{zhou2024easyjailbreak}.

These case studies collectively reveal the broad applicability of jailbreak prompts for malicious purposes and the necessity of a layered defense approach. By integrating prompt-level filtering, model-level mechanisms such as adversarial training and self-critique, and innovative strategies like unlearning harmful knowledge (e.g., {\it Eraser} method), we propose a comprehensive framework for mitigating these risks \cite{lu2024eraser}.

The findings presented in this paper aim to bridge the gap between the theoretical understanding of jailbreak prompts and practical defenses against their exploitation. By focusing on the cyber defense perspective, we aim to contribute to the development of secure, resilient, and ethically aligned AI systems that can withstand evolving threats in an increasingly interconnected world.

\section{Related work}

As LLMs progress in capability, adversarial techniques, including jailbreak attacks, have also become more sophisticated. Jailbreaking uses carefully crafted prompts to manipulate LLM outputs, bypassing built-in safeguards and revealing vulnerabilities in language generation \cite{liu2023autodan}. Research by \citet{liu2024hitchhiker} demonstrates that adversarial prompts can effectively circumvent safety measures and create risks that range from misinformation to ethical breaches.

\noindent
{\bf Categories and techniques of jailbreaking.} The techniques for jailbreak attacks can be divided into several main categories. \citet{cheng2024leveraging} introduced the {\it Contextual Interaction Attack}, leveraging multi-round interactions to progressively align model context with harmful intent. This approach builds a benign-seeming prompt context over multiple interactions, ultimately leading to harmful outputs. \citet{mehrotra2023tree} introduced the {\it Tree of Attacks with Pruning} (TAP), an iterative black-box method that improves jailbreak effectiveness while optimizing query efficiency by pruning ineffective prompts. Similarly, \citet{lu2024autojailbreak} presented {\it AutoJailbreak}, a framework that utilizes dependency analysis via directed acyclic graphs to position and examine jailbreak techniques in both attack and defense. Additionally, the {\it ObscurePrompt} method \cite{huang2024obscureprompt} explores the fragility of LLM alignments by manipulating out-of-distribution inputs to destabilize ethical boundaries, revealing overlooked vulnerabilities in alignment mechanisms.

\noindent
{\bf Evaluation and defense mechanisms.} Research has shown that jailbreak defenses face challenges in balancing model alignment and performance. For instance, \citet{liu2023autodan} proposed {\it AutoDAN}, an automated stealthy jailbreak attack using a genetic algorithm to bypass perplexity-based defenses by generating semantically coherent jailbreak prompts. \citet{pingua2024mitigating} developed {\it Prompt-G}, a defense mechanism that uses vector databases and embedding techniques to assess prompt safety in real time, effectively reducing attack success rates. Furthermore, {\it JailTrickBench} by \citet{xu2024bag} provides a standardized benchmarking approach to evaluate the efficacy of jailbreak defenses and emphasizes diverse testing environments and attack configurations. The introduction of {\it Defensive Prompt Patch (DPP)} showcases robust defenses against various jailbreak attacks while preserving the functional utility of LLMs \cite{xiong2024defensive}.

\noindent
{\bf Automated vs. handcrafted jailbreaks.} Handcrafted jailbreak prompts, such as {\it ``Do Anything Now (DAN),''} are widely recognized for their efficacy but face scalability issues. Automated frameworks, like {\it AutoJailbreak} \cite{lu2024autojailbreak} and {\it AutoDAN} \cite{liu2023autodan}, present a scalable alternative, achieving high transferability across LLMs without manual intervention. Frameworks like {\it WordGame} further extend attack methods by embedding adversarial queries into gamified prompts to evade traditional defenses \cite{zhang2024wordgame}. The {\it EasyJailbreak} unified framework provides modularity in constructing and evaluating various jailbreak methods, aiding both research and defense design \cite{zhou2024easyjailbreak}.

\noindent
{\bf Impact on security and ethical standards.} The ethical and security implications of jailbreak attacks are profound. \citet{liu2024hitchhiker} demonstrated that GPT-3.5 and GPT-4 could still generate harmful content under malicious prompts, despite robust filtering measures. Novel strategies like {\it Eraser} focus on unlearning harmful knowledge to mitigate security risks \cite{lu2024eraser}. These efforts underscore the need for continued development in dynamic defenses, balancing model robustness with minimizing ethical and operational risks \cite{pingua2024mitigating,lu2024autojailbreak}.

\section{Cyber defense perspective on preventing jailbreak prompts}

From a cyber defense perspective, preventing jailbreak prompts becomes essential to securing LLMs against misuse by cybercriminals. As LLMs integrate more deeply across sectors such as finance, healthcare, education, and critical infrastructure, their vulnerabilities to adversarial manipulation introduce risks that extend far beyond typical misuse scenarios. Cybercriminals exploit jailbreaks not only to generate phishing content, unauthorized code, and disinformation, but also to produce dangerously hazardous materials like bioweapons, explosive devices, and nuclear substances \cite{kirch2024features,zhang2024wordgame}. Moreover, recent research has demonstrated the effectiveness of advanced techniques such as {\it attention manipulation} (e.g., {\it AttnGCG}) to significantly improve the success rate of jailbreaks by exploiting the internal workings of transformer models \cite{wang2024attngcg}. These capabilities, when misappropriated, transform LLMs from tools of innovation into vectors for potentially catastrophic misuse, underscoring the urgency of a cyber defense strategy that can anticipate, prevent and adapt to these evolving threats.

Understanding jailbreak prompts as cyber threats shifts the focus toward robust adversarial defense frameworks, viewing jailbreak attempts as akin to other high-stakes cybersecurity threats, such as advanced persistent threats (APTs) or malware. Each jailbreak prompt can be seen as a targeted intrusion, deliberately designed to subvert model behaviors and generate information that would otherwise be restricted or ethically controlled \cite{liu2024hitchhiker}. Research by \citet{liu2023autodan} demonstrates that well-designed adversarial prompts can circumvent ethical boundaries in LLMs by exploiting context dependencies, semantic obfuscation, and even rhetorical ambiguity. Similarly, the vulnerabilities of LLMs to {\it multi-turn human jailbreaks}, which can achieve a success rate of over 70\% by exploiting conversational context, further underscore the limitations of current defense mechanisms and the need for comprehensive countermeasures \cite{li2024llm}. By categorizing jailbreak prompts according to their specific evasion techniques—such as single-turn attacks, context-building, prompt obfuscation, and dependency-driven manipulations—defenders can deploy tailored countermeasures against each category, optimizing LLM protection just as cybersecurity teams do for distinct types of digital threats.

A cyber defense perspective on LLM safety emphasizes the critical need for multi-layered defenses. This approach mirrors traditional cybersecurity principles, such as {\it ``defense in depth,''} which deploys multiple protective layers to mitigate risk even if one layer is compromised. In the context of LLMs, defenses can be applied across the entire workflow of prompt processing, response generation, and post-generation evaluation \cite{yi2024jailbreak}. {\bf Prompt-level defenses} include filters and detection algorithms that identify input patterns indicative of jailbreak attempts \cite{jain2023baseline,cao2023defending,robey2023smoothllm,sharma2024spml,zou2024system}. Recent advances also highlight strategies such as {\it defensive prompt patches} (DPP), which use suffix prompts to minimize attack success rates while preserving utility, making them a scalable solution for protecting LLMs \cite{xiong2024defensive}. For example, prompts that involve context-switching or obfuscation are flagged before they reach the model, helping to intercept adversarial inputs aimed at obtaining instructions for hazardous materials like bioweapons or bomb-making recipes. {\bf Model-level defenses} include self-critique mechanisms, ensemble evaluations, and adaptive response checks, which allow the model to evaluate its own outputs in real-time \cite{zeng2024autodefense,kim2024break,xie2024gradsafe,deng2023attack,bianchi2023safety,deng2023attack}. These model-level interventions act as a secondary layer of defense, refining outputs based on safety standards and preventing harmful responses if the prompt-level defense fails.

Adaptive learning is a critical element in this cyber defense framework, equipping LLM defenses to keep pace with rapidly evolving jailbreak techniques. Just as cybersecurity systems must constantly adapt to new forms of malware and exploits, LLM defenses must evolve to recognize new adversarial prompt patterns and counteract them effectively. For instance, using adversarial training, models can be exposed to simulated jailbreak prompts during their development phase, enabling them to better recognize and respond to real-world adversarial inputs once deployed. Furthermore, the development of systems like {\it Eraser}, which focus on unlearning harmful knowledge embedded in LLMs, prevents these models from producing dangerous outputs, thereby addressing the root cause of certain vulnerabilities \cite{lu2024eraser}. By creating a feedback loop in which LLMs continuously learn from attempted jailbreaks, defense systems can dynamically identify emerging patterns and incorporate these insights into ongoing model refinement. This approach parallels the use of honeypots and intrusion detection systems in cybersecurity, which monitor and analyze new threat patterns to improve resilience. As jailbreak methods evolve to bypass existing filters—such as through techniques like {\it AutoDAN}, which generates stealthy prompts that evade traditional perplexity-based detection—adaptive learning remains a crucial component of effective, long-term model security \cite{liu2023autodan}.

A defense-oriented approach also emphasizes monitoring, accountability, and transparency within LLM operations, drawing from best practices in cybersecurity governance. Just as cybersecurity frameworks include logging and auditing to detect and respond to unauthorized access, similar protocols are valuable in monitoring LLM interactions. Tracking and logging attempted jailbreaks, including successful or near-successful adversarial attempts, provide crucial data for improving defenses. These logs also support real-time anomaly detection, enabling prompt responses to suspicious activity. Furthermore, maintaining comprehensive records of LLM interactions enables organizations to conduct thorough audits, assess adherence to ethical standards, and refine defense protocols based on observed patterns. This transparency is crucial to foster trust, especially in high-stakes applications where LLM-generated outputs could be repurposed for illicit activities.

Beyond technical measures, the cyber defense perspective also incorporates regulatory and ethical safeguards that establish norms and guidelines for the use of LLM. Governments, industry groups, and regulatory bodies are increasingly recognizing the need for policies that govern LLM deployment, particularly in sensitive areas such as public safety, national security, and healthcare. Policies and standards that mandate security testing, incident reporting, and responsible use guidelines for LLMs help ensure that organizations prioritize ethical considerations alongside innovation. For instance, standards could require that LLM providers perform regular threat assessments, document potential misuse scenarios, and implement controls to prevent the generation of hazardous content like chemical weapon instructions. Collaboration across industries, academia, and government can foster shared security standards, enabling organizations to pool resources and knowledge for a more unified approach to combating adversarial threats against LLMs.

Understanding and addressing the threat of jailbreak prompts requires a proactive and layered approach. As adversaries continue to develop increasingly sophisticated methods to bypass ethical constraints and content filters, the challenge of securing LLMs against malicious misuse becomes more urgent. We examine case studies of jailbreak prompts to highlight the risks posed by cybercriminals exploiting LLM vulnerabilities. Each case emphasizes the adversarial techniques employed and the defense strategies needed to counter these evolving threats effectively.

\subsection{Case studies}\label{case_labels}

To better illustrate the diverse range of threats posed by jailbreak prompts and the corresponding defense strategies, this section presents six case studies. Each case study highlights a specific scenario where cybercriminals could exploit LLMs for malicious purposes, such as synthesizing bioweapons, bypassing nuclear safeguards, generating bomb-making instructions, automating financial fraud, spreading misinformation during public health crises, or hacking lottery systems. These examples underscore the critical need for robust, multi-layered defenses to mitigate the risks associated with the misuse of advanced language models. By analyzing these real-world-like scenarios, we aim to provide a comprehensive understanding of the challenges and propose practical solutions to counteract these threats effectively.

\noindent
{\bf Case study 1: Prevention of bioweapon instruction generation.} In this case study, we examine a scenario where a malicious actor attempts to exploit an LLM to obtain detailed instructions on synthesizing a bioweapon. This poses an extreme threat as biological agents can have devastating public health impacts. The defense strategy in this scenario would utilize a combination of prompt-level and model-level security layers. At the prompt level, keyword and phrase analysis would flag suspicious terms related to bioweapons, such as ``pathogen synthesis'' or ``biological warfare.'' Leveraging {\it Prompt-G}, an embedding-based system, the LLM would assess the context and intent of input prompts in real time, identifying both explicit and implied hazardous content \cite{pingua2024mitigating}. Additionally, methods like AttnGCG exploit vulnerabilities in attention mechanisms, highlighting the need for enhanced attention-weight monitoring to prevent adversarial manipulation \cite{wang2024attngcg}. 

On the model level, a self-critique mechanism would allow the LLM to evaluate its own outputs, checking for ethical alignment and content restrictions before finalizing responses. This layer could prevent unauthorized outputs from reaching the end-user even if the initial filters fail. In addition, monitoring functions would log suspicious prompts, allowing security teams to detect patterns and refine defenses continually. Logging also supports regulatory and legal accountability by creating a transparent audit trail, and enables rapid responses if the LLM is misused. This case underscores the importance of multi-layered defenses that work in tandem to detect and block highly sensitive prompts, protecting public safety from biohazard-related misuse.  

\noindent
{\bf Case study 2: Blocking nuclear material production queries.} This scenario addresses the potential misuse of LLMs to provide guidance on creating or acquiring nuclear materials. Given the catastrophic consequences of nuclear threats, this scenario requires advanced, adaptive learning mechanisms. The defense strategy here would involve not only prompt-level keyword detection but also the ability to recognize indirect or veiled queries on nuclear technology. For instance, prompts may use scientific or euphemistic language to disguise their intent, such as ``high-energy particle reactions'' or ``uranium processing.'' An adaptive learning system can evolve to detect these contextual disguises, using historical interaction data to recognize prompt patterns that may not explicitly refer to nuclear technology but imply hazardous intent \cite{liu2024hitchhiker}. Emerging methods like ObscurePrompt further complicate detection by disguising malicious queries in ambiguous text, necessitating improved syntax-agnostic filtration systems \cite{huang2024obscureprompt}. 

Incorporating {adversarial training}, where simulated jailbreak scenarios are applied during model development, enables the LLM to recognize and counter increasingly sophisticated prompt engineering tactics, especially in queries related to nuclear materials. This proactive approach mirrors the concept of red-teaming in cybersecurity, where defense strategies are tested and strengthened against realistic attack simulations. In addition, the inclusion of model-level ensemble responses, where multiple LLM versions independently assess and filter outputs, reinforces protection by making it harder for a single malicious prompt to bypass defenses. This case study highlights the importance of adapting model responses and training procedures to address sophisticated and evolving queries that aim to manipulate LLMs for sensitive and high-risk information.  

\noindent
{\bf Case study 3: Mitigating bomb-making instructions via contextual prompt analysis.} In this scenario, a cybercriminal employs iterative prompts in an attempt to elicit bomb-making instructions from the LLM. These prompts may start with innocuous-seeming questions about chemistry or mechanical processes, with each query gradually building toward the intended malicious output. The attacker exploits the multi-turn nature of conversations to lead the LLM down a path where, with context accumulation, it inadvertently provides the information needed to construct an explosive device. This incremental approach requires defenses capable of recognizing the broader context across multiple exchanges, as opposed to filtering each prompt in isolation \cite{cheng2024leveraging,liu2023autodan,zhang2024wordgame}. Research on multi-turn adversarial tactics such as those outlined in the Multi-Turn Human Jailbreak (MHJ) dataset reinforces the importance of tracking sequential prompts to prevent contextual manipulation \cite{li2024llm}. 

The defense strategy for this case relies heavily on contextual prompt analysis. Through sequential prompt analysis, the LLM tracks the progression of prompts and identifies when a series of seemingly benign questions begin to form a dangerous pattern. A layered defense model could integrate session-based tracking, monitoring the cumulative content across all prompts within a conversation. This approach ensures that context-switching prompts, which might evade isolated filters, are analyzed for potential malicious intent in a broader context. Moreover, using ensemble model responses enables each model version to independently review and verify the appropriateness of responses, mitigating the risk of a harmful output if one model fails to recognize the adversarial pattern.  

\noindent
{\bf Case study 4: Protecting financial systems from fraudulent transaction automation.} This case explores the risk that LLMs are used to automate fraudulent financial transactions, such as unauthorized transfers, by providing detailed instructions on bypassing banking security protocols or creating fake credentials. In this scenario, a cybercriminal could use jailbreak techniques to obtain step-by-step instructions from an LLM to commit fraud. Studies have shown that cybercriminals often exploit the ability of models to generate plausible financial advice, which can be weaponized if ethical boundaries are bypassed \cite{liu2023autodan,xu2024comprehensive}. Token-level attack methods like JailMine further exacerbate this risk by enabling precise, automated manipulations to exploit vulnerabilities in LLM responses, emphasizing the need for logit-based defensive strategies \cite{li2024lockpicking}. 

The defense strategy in this case would involve a multi-faceted approach, combining prompt-level keyword detection with a semantic analysis system that recognizes contextual intent related to banking fraud, such as ``payment reversal,'' ``unauthorized access,'' or ``fake credentials'' \cite{liu2023jailbreaking,kirch2024features}. Research by \citet{xu2024comprehensive} emphasize that adaptive training that includes simulated fraudulent prompts, aligned with evolving attack tactics in financial sectors, is essential to improve the resilience of the model against adversarial manipulation. Model-level self-critique mechanisms would also be crucial, enabling the LLM to internally evaluate and flag responses that may violate financial security standards before output. In addition, logging interactions containing potentially harmful financial terms and phrases supports real-time monitoring and auditing, which can help refine detection and ensure regulatory compliance. This use case illustrates the need for continuous adaptation to protect sensitive sectors like finance, where unauthorized access or instructions to bypass security could lead to financial losses and harm to individuals and institutions.  

\noindent
{\bf Case study 5: Preventing the dissemination of misinformation related to public health emergencies.} In this case study, a cybercriminal might attempt to leverage an LLM to generate misinformation regarding public health emergencies, such as pandemics or outbreaks, intending to sow panic, disrupt social stability, or spread harmful health advice. Research has documented the potential for LLMs to inadvertently generate health-related misinformation if ethical filters are bypassed or manipulated \cite{xu2024comprehensive, liu2023autodan}. Through jailbreak prompts, the adversary could manipulate the LLM into creating false information or conspiracy theories about vaccines, treatments, or preventive measures, which can lead to public distrust and harmful behaviors. 

To address this, the cyber defense framework would employ context-aware prompt filtering that detects keywords associated with medical advice and public health emergencies. Real-time intent analysis would flag prompts that contain misinformation or advice that deviate from recognized health guidelines \cite{liu2024hitchhiker}. The integration of unlearning-based strategies, as demonstrated by methods like {\it Eraser}, ensures that harmful knowledge is systematically removed, significantly reducing the risk of LLM misuse in such contexts \cite{lu2024eraser}. \citet{xu2024comprehensive} recommend ensemble model evaluations for public health-related outputs, where outputs are cross-verified by multiple model versions to ensure consistency and adherence to safety protocols. Incorporating an ethics and misinformation detection layer could further improve resilience, enabling LLMs to recognize and counter common health-related falsehoods through adaptive learning. This aligns with the findings of \citet{xu2024comprehensive}, where context-sensitive multilayered defenses are shown to mitigate the risk of misuse in high-stakes areas such as public health, preserving public trust during crises.

\noindent
{\bf Case study 6: Exploiting LLMs for hacking lottery systems.}  
In this case study, a cybercriminal aims to use an LLM to identify vulnerabilities within a digital lottery system. By leveraging jailbreak prompts, the attacker could attempt to gain detailed knowledge of specific lottery system hacking techniques, such as manipulating random number generators, intercepting prize allocation algorithms, or bypassing authentication mechanisms. Studies have shown that adversarial prompts can effectively coerce LLMs into providing technically sophisticated guidance on bypassing security protocols if ethical filters are bypassed \cite{pa2023attacker,guo2024cold,abdali2024securing,pa2023attacker}. In this scenario, the attacker might query the LLM with questions framed as ``hypothetical scenarios'' to circumvent filters, gradually escalating the prompts to request more specific details about vulnerabilities in lottery algorithms \cite{yi2024jailbreak,kirch2024features}. Emerging approaches such as {\it EasyJailbreak} reveal modular frameworks that attackers could exploit to test various combinations of adversarial strategies, further complicating detection and response efforts \cite{zhou2024easyjailbreak}.

A defense strategy for this type of misuse involves both prompt-level filtering and advanced intent analysis to recognize subtle, context-driven attempts to gather hacking instructions. At the model level, ensemble evaluations and adversarial training could help the LLM identify and block prompts that combine seemingly benign questions into harmful guidance \cite{chen2024characterizing}. Real-time logging and pattern recognition tools would further support this defense by detecting interactions related to specific hacking terminology and flagging them for immediate review \cite{yu2024don,li2024llm}. By employing layered, context-sensitive defenses, lottery systems, and related organizations can reduce the risk of their security being compromised through adversarial use of LLMs, protecting both the integrity of the system and the trust of its participants. 

\section{Discussion}
The case studies presented in this work highlight the critical importance of addressing jailbreak prompts as malicious tools leveraged by cybercriminals. These adversarial strategies exploit vulnerabilities in LLMs, posing significant challenges to their safe and ethical deployment. Each scenario—from bioweapon synthesis and financial fraud to misinformation dissemination and lottery system hacking—demonstrates the diverse ways cybercriminals can misuse LLMs, emphasizing the need for robust, multi-faceted defenses to safeguard against these threats.

One key observation is the escalating sophistication of adversarial techniques, such as multi-turn jailbreaks and token-level manipulations, which exploit weaknesses in existing defenses. For instance, the MHJ dataset reveals a significant gap in current LLM defenses, achieving high attack success rates through sequential prompt manipulation \cite{li2024llm}. Similarly, methods like {\it AttnGCG} and {\it ObscurePrompt} expose vulnerabilities in attention mechanisms and out-of-distribution prompt handling, necessitating advanced syntax-agnostic filtration systems \cite{wang2024attngcg,huang2024obscureprompt}.

The trade-off between safety and utility remains a central challenge. Defensive measures such as adversarial training and ensemble evaluations, while effective, may inadvertently reduce the functionality of LLMs for legitimate use cases. For example, unlearning-based methods like {\it Eraser} show promise in mitigating residual harmful knowledge but require careful calibration to maintain general-purpose capabilities \cite{lu2024eraser}. Additionally, the computational cost of real-time defenses, such as logit-based monitoring in token-level attacks, highlights the need for scalable and efficient solutions \cite{li2024lockpicking}.

Layered defense frameworks emerge as a critical approach. Combining prompt-level filtering, model-level self-critique mechanisms, and context-aware tracking ensures a robust response to both explicit and implicit adversarial prompts. Tools like {\it EasyJailbreak} offer modular frameworks for testing and improving these layered defenses, underscoring the importance of collaboration and open research in combating cyber threats \cite{zhou2024easyjailbreak}.

Finally, fostering transparency and ethical governance in LLM deployment is imperative. Sharing datasets, red-teaming insights, and defense strategies can empower stakeholders to anticipate and mitigate emerging threats effectively. As demonstrated by the MHJ dataset and JailbreakHub framework, collective efforts are essential to align LLM development with societal safety standards \cite{li2024llm,shen2023anything}.

\section{Conclusion}

This paper emphasizes the critical importance of preventing jailbreak prompts from being exploited as malicious tools by cybercriminals. Through an in-depth analysis of diverse case studies, we have demonstrated the vulnerabilities of LLMs to adversarial manipulation and outlined multi-layered defense strategies to address these risks. These strategies include prompt-level filtering, adversarial training, model self-critique, and unlearning harmful knowledge, each contributing to a robust cyber defense framework.

The findings highlight that defending against jailbreak prompts requires continuous innovation, collaboration, and ethical oversight. The sophistication of attacks, such as multi-turn jailbreaks and logit-based token manipulation, underscores the need for adaptive and scalable defenses. Moreover, balancing safety and utility remains a pressing challenge that demands careful calibration of LLM systems.

As LLMs continue to revolutionize industries, their security must remain a top priority. By fostering transparency, open research, and ethical principles, the community can mitigate the misuse of LLMs and ensure their responsible deployment. Ultimately, safeguarding these transformative technologies against cybercriminal exploitation will enable their potential to be fully realized in a manner aligned with societal values and safety.


\bibliography{latex/custom.bib}

\begin{thebibliography}{47}
\providecommand{\natexlab}[1]{#1}

\bibitem[{Abdali et~al.(2024)Abdali, Anarfi, Barberan, and He}]{abdali2024securing}
Sara Abdali, Richard Anarfi, CJ~Barberan, and Jia He. 2024.
\newblock Securing large language models: Threats, vulnerabilities and responsible practices.
\newblock \emph{arXiv preprint arXiv:2403.12503}.

\bibitem[{Achiam et~al.(2023)Achiam, Adler, Agarwal, Ahmad, Akkaya, Aleman, Almeida, Altenschmidt, Altman, Anadkat et~al.}]{achiam2023gpt}
Josh Achiam, Steven Adler, Sandhini Agarwal, Lama Ahmad, Ilge Akkaya, Florencia~Leoni Aleman, Diogo Almeida, Janko Altenschmidt, Sam Altman, Shyamal Anadkat, et~al. 2023.
\newblock Gpt-4 technical report.
\newblock \emph{arXiv preprint arXiv:2303.08774}.

\bibitem[{Andriushchenko et~al.(2024)Andriushchenko, Croce, and Flammarion}]{andriushchenko2024jailbreaking}
Maksym Andriushchenko, Francesco Croce, and Nicolas Flammarion. 2024.
\newblock Jailbreaking leading safety-aligned llms with simple adaptive attacks.
\newblock \emph{arXiv preprint arXiv:2404.02151}.

\bibitem[{Bianchi et~al.(2023)Bianchi, Suzgun, Attanasio, R{\"o}ttger, Jurafsky, Hashimoto, and Zou}]{bianchi2023safety}
Federico Bianchi, Mirac Suzgun, Giuseppe Attanasio, Paul R{\"o}ttger, Dan Jurafsky, Tatsunori Hashimoto, and James Zou. 2023.
\newblock Safety-tuned llamas: Lessons from improving the safety of large language models that follow instructions.
\newblock \emph{arXiv preprint arXiv:2309.07875}.

\bibitem[{Cao et~al.(2023)Cao, Cao, Lin, and Chen}]{cao2023defending}
Bochuan Cao, Yuanpu Cao, Lu~Lin, and Jinghui Chen. 2023.
\newblock Defending against alignment-breaking attacks via robustly aligned llm.
\newblock \emph{arXiv preprint arXiv:2309.14348}.

\bibitem[{Chen et~al.(2024)Chen, Liu, Wang, Chen, and Wang}]{chen2024characterizing}
Kexin Chen, Yi~Liu, Dongxia Wang, Jiaying Chen, and Wenhai Wang. 2024.
\newblock Characterizing and evaluating the reliability of llms against jailbreak attacks.
\newblock \emph{arXiv preprint arXiv:2408.09326}.

\bibitem[{Cheng et~al.(2024)Cheng, Georgopoulos, Cevher, and Chrysos}]{cheng2024leveraging}
Yixin Cheng, Markos Georgopoulos, Volkan Cevher, and Grigorios~G Chrysos. 2024.
\newblock Leveraging the context through multi-round interactions for jailbreaking attacks.
\newblock \emph{arXiv preprint arXiv:2402.09177}.

\bibitem[{Deng et~al.(2023{\natexlab{a}})Deng, Wang, Feng, Deng, Wang, and He}]{deng2023attack}
Boyi Deng, Wenjie Wang, Fuli Feng, Yang Deng, Qifan Wang, and Xiangnan He. 2023{\natexlab{a}}.
\newblock Attack prompt generation for red teaming and defending large language models.
\newblock \emph{arXiv preprint arXiv:2310.12505}.

\bibitem[{Deng et~al.(2023{\natexlab{b}})Deng, Liu, Li, Wang, Zhang, Li, Wang, Zhang, and Liu}]{deng2023jailbreaker}
Gelei Deng, Yi~Liu, Yuekang Li, Kailong Wang, Ying Zhang, Zefeng Li, Haoyu Wang, Tianwei Zhang, and Yang Liu. 2023{\natexlab{b}}.
\newblock Jailbreaker: Automated jailbreak across multiple large language model chatbots.
\newblock \emph{arXiv preprint arXiv:2307.08715}.

\bibitem[{Guo et~al.(2024)Guo, Yu, Zhang, Qin, and Hu}]{guo2024cold}
Xingang Guo, Fangxu Yu, Huan Zhang, Lianhui Qin, and Bin Hu. 2024.
\newblock Cold-attack: Jailbreaking llms with stealthiness and controllability.
\newblock \emph{arXiv preprint arXiv:2402.08679}.

\bibitem[{Gupta et~al.(2023)Gupta, Akiri, Aryal, Parker, and Praharaj}]{gupta2023chatgpt}
Maanak Gupta, CharanKumar Akiri, Kshitiz Aryal, Eli Parker, and Lopamudra Praharaj. 2023.
\newblock From chatgpt to threatgpt: Impact of generative ai in cybersecurity and privacy.
\newblock \emph{IEEE Access}.

\bibitem[{Huang et~al.(2024)Huang, Tang, Chen, Tang, Wan, Sun, and Zhang}]{huang2024obscureprompt}
Yue Huang, Jingyu Tang, Dongping Chen, Bingda Tang, Yao Wan, Lichao Sun, and Xiangliang Zhang. 2024.
\newblock Obscureprompt: Jailbreaking large language models via obscure input.
\newblock \emph{arXiv preprint arXiv:2406.13662}.

\bibitem[{Jain et~al.(2023)Jain, Schwarzschild, Wen, Somepalli, Kirchenbauer, Chiang, Goldblum, Saha, Geiping, and Goldstein}]{jain2023baseline}
Neel Jain, Avi Schwarzschild, Yuxin Wen, Gowthami Somepalli, John Kirchenbauer, Ping-yeh Chiang, Micah Goldblum, Aniruddha Saha, Jonas Geiping, and Tom Goldstein. 2023.
\newblock Baseline defenses for adversarial attacks against aligned language models.
\newblock \emph{arXiv preprint arXiv:2309.00614}.

\bibitem[{Kim et~al.(2024)Kim, Yuk, and Cho}]{kim2024break}
Heegyu Kim, Sehyun Yuk, and Hyunsouk Cho. 2024.
\newblock Break the breakout: Reinventing lm defense against jailbreak attacks with self-refinement.
\newblock \emph{arXiv preprint arXiv:2402.15180}.

\bibitem[{Kirch et~al.(2024)Kirch, Field, and Casper}]{kirch2024features}
Nathalie~Maria Kirch, Severin Field, and Stephen Casper. 2024.
\newblock What features in prompts jailbreak llms? investigating the mechanisms behind attacks.
\newblock \emph{arXiv preprint arXiv:2411.03343}.

\bibitem[{Li et~al.(2024{\natexlab{a}})Li, Han, Steneker, Primack, Goodside, Zhang, Wang, Menghini, and Yue}]{li2024llm}
Nathaniel Li, Ziwen Han, Ian Steneker, Willow Primack, Riley Goodside, Hugh Zhang, Zifan Wang, Cristina Menghini, and Summer Yue. 2024{\natexlab{a}}.
\newblock Llm defenses are not robust to multi-turn human jailbreaks yet.
\newblock \emph{arXiv preprint arXiv:2408.15221}.

\bibitem[{Li et~al.(2024{\natexlab{b}})Li, Wang, Cheng, Zhou, and Hsieh}]{li2024drattack}
Xirui Li, Ruochen Wang, Minhao Cheng, Tianyi Zhou, and Cho-Jui Hsieh. 2024{\natexlab{b}}.
\newblock Drattack: Prompt decomposition and reconstruction makes powerful llm jailbreakers.
\newblock \emph{arXiv preprint arXiv:2402.16914}.

\bibitem[{Li et~al.(2024{\natexlab{c}})Li, Liu, Li, Shi, Deng, Chen, and Wang}]{li2024lockpicking}
Yuxi Li, Yi~Liu, Yuekang Li, Ling Shi, Gelei Deng, Shengquan Chen, and Kailong Wang. 2024{\natexlab{c}}.
\newblock Lockpicking llms: A logit-based jailbreak using token-level manipulation.
\newblock \emph{arXiv preprint arXiv:2405.13068}.

\bibitem[{Liu et~al.(2023{\natexlab{a}})Liu, Xu, Chen, and Xiao}]{liu2023autodan}
Xiaogeng Liu, Nan Xu, Muhao Chen, and Chaowei Xiao. 2023{\natexlab{a}}.
\newblock Autodan: Generating stealthy jailbreak prompts on aligned large language models.
\newblock \emph{arXiv preprint arXiv:2310.04451}.

\bibitem[{Liu et~al.(2024)Liu, Deng, Xu, Li, Zheng, Zhang, Zhao, Zhang, and Wang}]{liu2024hitchhiker}
Yi~Liu, Gelei Deng, Zhengzi Xu, Yuekang Li, Yaowen Zheng, Ying Zhang, Lida Zhao, Tianwei Zhang, and Kailong Wang. 2024.
\newblock A hitchhiker’s guide to jailbreaking chatgpt via prompt engineering.
\newblock In \emph{Proceedings of the 4th International Workshop on Software Engineering and AI for Data Quality in Cyber-Physical Systems/Internet of Things}, pages 12--21.

\bibitem[{Liu et~al.(2023{\natexlab{b}})Liu, Deng, Xu, Li, Zheng, Zhang, Zhao, Zhang, Wang, and Liu}]{liu2023jailbreaking}
Yi~Liu, Gelei Deng, Zhengzi Xu, Yuekang Li, Yaowen Zheng, Ying Zhang, Lida Zhao, Tianwei Zhang, Kailong Wang, and Yang Liu. 2023{\natexlab{b}}.
\newblock Jailbreaking chatgpt via prompt engineering: An empirical study.
\newblock \emph{arXiv preprint arXiv:2305.13860}.

\bibitem[{Lu et~al.(2024{\natexlab{a}})Lu, Yan, Yuan, Shi, Wei, Chen, and Zhou}]{lu2024autojailbreak}
Lin Lu, Hai Yan, Zenghui Yuan, Jiawen Shi, Wenqi Wei, Pin-Yu Chen, and Pan Zhou. 2024{\natexlab{a}}.
\newblock Autojailbreak: Exploring jailbreak attacks and defenses through a dependency lens.
\newblock \emph{arXiv preprint arXiv:2406.03805}.

\bibitem[{Lu et~al.(2024{\natexlab{b}})Lu, Zeng, Wang, Lu, Chen, Zhuang, and Chen}]{lu2024eraser}
Weikai Lu, Ziqian Zeng, Jianwei Wang, Zhengdong Lu, Zelin Chen, Huiping Zhuang, and Cen Chen. 2024{\natexlab{b}}.
\newblock Eraser: Jailbreaking defense in large language models via unlearning harmful knowledge.
\newblock \emph{arXiv preprint arXiv:2404.05880}.

\bibitem[{Luo et~al.(2024)Luo, Ma, Liu, Guo, and Xiao}]{luo2024jailbreakv}
Weidi Luo, Siyuan Ma, Xiaogeng Liu, Xiaoyu Guo, and Chaowei Xiao. 2024.
\newblock Jailbreakv-28k: A benchmark for assessing the robustness of multimodal large language models against jailbreak attacks.
\newblock \emph{arXiv preprint arXiv:2404.03027}.

\bibitem[{Mehrotra et~al.(2023)Mehrotra, Zampetakis, Kassianik, Nelson, Anderson, Singer, and Karbasi}]{mehrotra2023tree}
Anay Mehrotra, Manolis Zampetakis, Paul Kassianik, Blaine Nelson, Hyrum Anderson, Yaron Singer, and Amin Karbasi. 2023.
\newblock Tree of attacks: Jailbreaking black-box llms automatically.
\newblock \emph{arXiv preprint arXiv:2312.02119}.

\bibitem[{Pa~Pa et~al.(2023)Pa~Pa, Tanizaki, Kou, Van~Eeten, Yoshioka, and Matsumoto}]{pa2023attacker}
Yin~Minn Pa~Pa, Shunsuke Tanizaki, Tetsui Kou, Michel Van~Eeten, Katsunari Yoshioka, and Tsutomu Matsumoto. 2023.
\newblock An attacker’s dream? exploring the capabilities of chatgpt for developing malware.
\newblock In \emph{Proceedings of the 16th Cyber Security Experimentation and Test Workshop}, pages 10--18.

\bibitem[{Peng et~al.(2024)Peng, Bi, Niu, Liu, Feng, Wang, Yan, Wen, Zhang, and Yin}]{peng2024jailbreaking}
Benji Peng, Ziqian Bi, Qian Niu, Ming Liu, Pohsun Feng, Tianyang Wang, Lawrence~KQ Yan, Yizhu Wen, Yichao Zhang, and Caitlyn~Heqi Yin. 2024.
\newblock Jailbreaking and mitigation of vulnerabilities in large language models.
\newblock \emph{arXiv preprint arXiv:2410.15236}.

\bibitem[{Pingua et~al.(2024)Pingua, Murmu, Kandpal, Rautaray, Mishra, Barik, and Saikia}]{pingua2024mitigating}
Bhagyajit Pingua, Deepak Murmu, Meenakshi Kandpal, Jyotirmayee Rautaray, Pranati Mishra, Rabindra~Kumar Barik, and Manob~Jyoti Saikia. 2024.
\newblock Mitigating adversarial manipulation in llms: a prompt-based approach to counter jailbreak attacks (prompt-g).
\newblock \emph{PeerJ Computer Science}, 10:e2374.

\bibitem[{Robey et~al.(2023)Robey, Wong, Hassani, and Pappas}]{robey2023smoothllm}
Alexander Robey, Eric Wong, Hamed Hassani, and George~J Pappas. 2023.
\newblock Smoothllm: Defending large language models against jailbreaking attacks.
\newblock \emph{arXiv preprint arXiv:2310.03684}.

\bibitem[{Sharma et~al.(2024)Sharma, Gupta, and Grossman}]{sharma2024spml}
Reshabh~K Sharma, Vinayak Gupta, and Dan Grossman. 2024.
\newblock Spml: A dsl for defending language models against prompt attacks.
\newblock \emph{arXiv preprint arXiv:2402.11755}.

\bibitem[{Shen et~al.(2023)Shen, Chen, Backes, Shen, and Zhang}]{shen2023anything}
Xinyue Shen, Zeyuan Chen, Michael Backes, Yun Shen, and Yang Zhang. 2023.
\newblock " do anything now": Characterizing and evaluating in-the-wild jailbreak prompts on large language models.
\newblock \emph{arXiv preprint arXiv:2308.03825}.

\bibitem[{Wang et~al.(2024)Wang, Tu, Mei, Zhao, Wang, and Xie}]{wang2024attngcg}
Zijun Wang, Haoqin Tu, Jieru Mei, Bingchen Zhao, Yisen Wang, and Cihang Xie. 2024.
\newblock Attngcg: Enhancing jailbreaking attacks on llms with attention manipulation.
\newblock \emph{arXiv preprint arXiv:2410.09040}.

\bibitem[{Wei et~al.(2024)Wei, Haghtalab, and Steinhardt}]{wei2024jailbroken}
Alexander Wei, Nika Haghtalab, and Jacob Steinhardt. 2024.
\newblock Jailbroken: How does llm safety training fail?
\newblock \emph{Advances in Neural Information Processing Systems}, 36.

\bibitem[{Xie et~al.(2024)Xie, Fang, Pi, and Gong}]{xie2024gradsafe}
Yueqi Xie, Minghong Fang, Renjie Pi, and Neil Gong. 2024.
\newblock Gradsafe: Detecting unsafe prompts for llms via safety-critical gradient analysis.
\newblock \emph{arXiv preprint arXiv:2402.13494}.

\bibitem[{Xie et~al.(2023)Xie, Yi, Shao, Curl, Lyu, Chen, Xie, and Wu}]{xie2023defending}
Yueqi Xie, Jingwei Yi, Jiawei Shao, Justin Curl, Lingjuan Lyu, Qifeng Chen, Xing Xie, and Fangzhao Wu. 2023.
\newblock Defending chatgpt against jailbreak attack via self-reminders.
\newblock \emph{Nature Machine Intelligence}, 5(12):1486--1496.

\bibitem[{Xiong et~al.(2024)Xiong, Qi, Chen, and Ho}]{xiong2024defensive}
Chen Xiong, Xiangyu Qi, Pin-Yu Chen, and Tsung-Yi Ho. 2024.
\newblock Defensive prompt patch: A robust and interpretable defense of llms against jailbreak attacks.
\newblock \emph{arXiv preprint arXiv:2405.20099}.

\bibitem[{Xu et~al.(2024{\natexlab{a}})Xu, Stokes, McDonald, Bai, Marshall, Wang, Swaminathan, and Li}]{xu2024autoattacker}
Jiacen Xu, Jack~W Stokes, Geoff McDonald, Xuesong Bai, David Marshall, Siyue Wang, Adith Swaminathan, and Zhou Li. 2024{\natexlab{a}}.
\newblock Autoattacker: A large language model guided system to implement automatic cyber-attacks.
\newblock \emph{arXiv preprint arXiv:2403.01038}.

\bibitem[{Xu et~al.(2024{\natexlab{b}})Xu, Liu, and Liu}]{xu2024bag}
Zhao Xu, Fan Liu, and Hao Liu. 2024{\natexlab{b}}.
\newblock Bag of tricks: Benchmarking of jailbreak attacks on llms.
\newblock \emph{arXiv preprint arXiv:2406.09324}.

\bibitem[{Xu et~al.(2024{\natexlab{c}})Xu, Liu, Deng, Li, and Picek}]{xu2024comprehensive}
Zihao Xu, Yi~Liu, Gelei Deng, Yuekang Li, and Stjepan Picek. 2024{\natexlab{c}}.
\newblock A comprehensive study of jailbreak attack versus defense for large language models.
\newblock In \emph{Findings of the Association for Computational Linguistics ACL 2024}, pages 7432--7449.

\bibitem[{Xu et~al.(2024{\natexlab{d}})Xu, Liu, Deng, Li, and Picek}]{xu2024llm}
Zihao Xu, Yi~Liu, Gelei Deng, Yuekang Li, and Stjepan Picek. 2024{\natexlab{d}}.
\newblock Llm jailbreak attack versus defense techniques--a comprehensive study.
\newblock \emph{arXiv preprint arXiv:2402.13457}.

\bibitem[{Yi et~al.(2024)Yi, Liu, Sun, Cong, He, Song, Xu, and Li}]{yi2024jailbreak}
Sibo Yi, Yule Liu, Zhen Sun, Tianshuo Cong, Xinlei He, Jiaxing Song, Ke~Xu, and Qi~Li. 2024.
\newblock Jailbreak attacks and defenses against large language models: A survey.
\newblock \emph{arXiv preprint arXiv:2407.04295}.

\bibitem[{Yu et~al.(2024)Yu, Liu, Liang, Cameron, Xiao, and Zhang}]{yu2024don}
Zhiyuan Yu, Xiaogeng Liu, Shunning Liang, Zach Cameron, Chaowei Xiao, and Ning Zhang. 2024.
\newblock Don't listen to me: Understanding and exploring jailbreak prompts of large language models.
\newblock \emph{arXiv preprint arXiv:2403.17336}.

\bibitem[{Zeng et~al.(2024)Zeng, Wu, Zhang, Wang, and Wu}]{zeng2024autodefense}
Yifan Zeng, Yiran Wu, Xiao Zhang, Huazheng Wang, and Qingyun Wu. 2024.
\newblock Autodefense: Multi-agent llm defense against jailbreak attacks.
\newblock \emph{arXiv preprint arXiv:2403.04783}.

\bibitem[{Zhang et~al.(2024{\natexlab{a}})Zhang, Bu, Wen, Chen, Li, and Zhu}]{zhang2024llms}
Jie Zhang, Haoyu Bu, Hui Wen, Yu~Chen, Lun Li, and Hongsong Zhu. 2024{\natexlab{a}}.
\newblock When llms meet cybersecurity: A systematic literature review.
\newblock \emph{arXiv preprint arXiv:2405.03644}.

\bibitem[{Zhang et~al.(2024{\natexlab{b}})Zhang, Cao, Cao, Lin, Mitra, and Chen}]{zhang2024wordgame}
Tianrong Zhang, Bochuan Cao, Yuanpu Cao, Lu~Lin, Prasenjit Mitra, and Jinghui Chen. 2024{\natexlab{b}}.
\newblock Wordgame: Efficient \& effective llm jailbreak via simultaneous obfuscation in query and response.
\newblock \emph{arXiv preprint arXiv:2405.14023}.

\bibitem[{Zhou et~al.(2024)Zhou, Wang, Xiong, Xia, Gu, Chai, Zhu, Huang, Dou, Xi et~al.}]{zhou2024easyjailbreak}
Weikang Zhou, Xiao Wang, Limao Xiong, Han Xia, Yingshuang Gu, Mingxu Chai, Fukang Zhu, Caishuang Huang, Shihan Dou, Zhiheng Xi, et~al. 2024.
\newblock Easyjailbreak: A unified framework for jailbreaking large language models.
\newblock \emph{arXiv preprint arXiv:2403.12171}.

\bibitem[{Zou et~al.(2024)Zou, Chen, and Li}]{zou2024system}
Xiaotian Zou, Yongkang Chen, and Ke~Li. 2024.
\newblock Is the system message really important to jailbreaks in large language models?
\newblock \emph{arXiv preprint arXiv:2402.14857}.

\end{thebibliography}

\end{document}